\documentclass[twocolumn,showpacs,preprintnumbers,amsmath,amssymb]{revtex4}
\usepackage{amssymb}%
\usepackage{graphicx}
\usepackage{dcolumn}
\usepackage{bm}

\newcommand{\bi}{\begin{itemize}}
\newcommand{\ei}{\end{itemize}}
\newcommand{\be}{\begin{equation}}
\newcommand{\ee}{\end{equation}}
\newcommand{\ba}{\begin{eqnarray}}
\newcommand{\ea}{\end{eqnarray}}
\newcommand{\bse}{\begin{subequations}}
\newcommand{\ese}{\end{subequations}}

\newcommand{\tbr}{t_{\textrm{\tiny{2BR}}}}

\begin{document}

\input{epsf.sty}

\title{Scalar field ``mini--MACHOs'': a new explanation for 
galactic dark matter}

\author{Xavier Hern\'andez$^{1}$, Tonatiuh~Matos$^{2}$, Roberto A
Sussman$^{3}$ and Yosef Verbin$^4$}
\affiliation{$^{1}$Instituto de Astronom\'\i a - UNAM, Apdo. 70-267,
Ciudad
Universitaria, 04510 M\'exico, D.F., M\'exico.\\$^{2}$ Departamento de
F{\'\i}sica,
Centro de Investigaci\'on y de Estudios Avanzados del IPN, A.P. 14-740,
07000 M\'exico D.F., M\'exico.\\$^{3}$Instituto de Ciencias
Nucleares - UNAM, Apdo. 70-543, Ciudad Universitaria, 04510
M\'exico, D.F., M\'exico.\\$^4$ Department of Natural Sciences, The Open
University of Israel, P.O.B. 39328, Tel Aviv 61392, Israel.}

\email{xavier@astroscu.unam.mx, tmatos@fis.cinvestav.mx,
sussman@nuclecu.unam.mx, verbin@oumail.openu.ac.il}


\begin{abstract}
We examine the possibility that galactic halos are collisionless ensembles
of scalar field ``massive compact halo objects'' (MACHOs). Using mass constraints
from MACHO microlensing and from theoretical arguments on halos made up of massive
black holes, as well as demanding also that scalar MACHO ensambles of all
scales do not exhibit gravothermal instability (as required by consistency with
observations of LSB galaxies), we obtain the range:\, $m\alt 10^{-7}\, M_\odot$
\,or\, $30\, M_\odot\alt m\alt 100\,M_\odot$. The rather narrow mass range of large
MACHOs seems to indicate that the ensambles we are suggesting should be probably
made up of scalar MACHOs in the low mass range (``mini--MACHOs''). The proposed
model allows one to consider a non--baryonic and non--thermal fundamental nature
of dark matter, while at the same time keeping the same phenomenology of the CDM
paradigm.
\end{abstract}


\maketitle

\section{Introduction}

The problem of the missing mass on galactic and galactic
cluster scales is one of the most interesting open issues in present
day cosmology and astrophysics~\cite{DMreview}. At cosmological scales, recent
WMAP results~\cite{MAP} have confirmed that the global  dynamics of the universe
imply a far larger non relativistic "matter"  component than what is allowed by
light density and nucleosynthesis  estimates of baryonic matter, further
strengthening the conclusion of a significant "cold dark matter" contribution at
all scales. The dominant approach to this problem has been the so-called ``cold
dark matter'' (CDM) paradigm in which the missing mass-energy in galactic halos
is made of relic gases of new and yet undetected
types of elementary non--baryonic (possibly supersymmetric) particles, collectively
known as weakly interacting massive particles (WIMPs).

Since all theories unifying gravity with other interactions
involve scalar fields, more ``exotic'' scenarios consider dark
matter in the form of a scalar field coherent on a very large
scale, similar to those associated with quintessence sources.
Kaluza--Klein, Super strings theories and super-gravity
\cite{Kallosh}, all contain scalar fields as reminiscent of extra dimension
of spacetime. Even if, until now, these remnants of primordial scalar
fields have not been directly detected, their use as models for dark energy
\cite{steinhardt} or dark matter in galactic halo structures
\cite{GuzmanMatos2000} has become widespread.

So far, most of the attempts to model galactic dark matter halos
out of real or complex scalar fields assume that each galactic
halo is a spherical Bose-Einstein condensate of scalar particles.
This was first suggested \cite{SJSin} assuming free ultra light
($\sim10^{-24}$eV) scalar particles described by a coherent
complex scalar field forming a ``boson star'' of a galactic scale.
Subsequent studies \cite{LeeKoh} added self-interaction and
generalized the previous Newtonian analysis to be fully general
relativistic. Other authors contributed more detailed studies
which used two kinds of coupling of the scalar field to gravity,
i.e. either minimal
\cite{MatosEtAl2000,GuzmanMatos2000,MatosGuzman2001, Arbey,bosonS} or
non-minimal\cite{ NucamendiEtAl2000,NucamendiEtAl2001}. Static
solutions (``boson stars'') are possible with a complex field, but
not for a real valued field. The latter do allow for stable
oscillating objects called ``oscillatons'' that can be used to
model galactic structures of all known scales (see
\cite{seidel,chipocludo,SFDM}).

\section{Scalar fields halo models} 

Both, oscillatons and boson stars, are described by the field theoretical action
\begin{eqnarray}
S=\int d^{4} x\sqrt{|g|}\left( \frac{1}{2}(\nabla_\mu
\Phi)^*(\nabla ^\mu \Phi)-U(|\Phi|)+\frac{{\cal R}}{16\pi \,G}\right)
\label{action1}
\end{eqnarray}
where $g,\,{\cal R}$ are the metric determinant and Ricci scalar
and $\Phi$ is a real or complex scalar field, which will give
rise, respectively, to oscillatons and boson stars. We look at
each case separately below.

An example of stable oscillatons \cite{seidel} is examined in
\cite{chipocludo,luis,phi2,LuisPaco}, corresponding to scalar field with
mass $m_{\Phi}$ and potential $U = m_{\Phi}^2\Phi^2/2$, which
forms stable objects with a critical mass given by
\begin{equation}
m_{\textrm{crit}} \ = \ 0.607 \ \frac{m_{Pl}^2}{m_{\Phi}}
\label{Mcrit}
\end{equation}
where $m_{Pl}$ is Planck's mass. Since we do not have any
criterion for choosing the scalar field mass, this parameter must
be selected by demanding that $m\simeq  m_{\textrm{crit}}$
complies with appropriate ranges. The formation of the
oscillaton depends on initial conditions set after inflation, when
the scalar field quantum fluctuations grow very fast and form seed
fluctuations with different sizes. If the seed fluctuation is
small, the oscillatons virialize and form a compact object in a
short time after inflation. If the seed fluctuation is large but
smaller than the critical mass, this virialization process (not to
confuse with that of the MACHO ensemble) takes longer, while if
the seed fluctuation is larger than the critical mass, the
oscillaton is no longer stable. However, for masses below the
critical mass the oscillatons are very stable
objects\cite{chipocludo}, even long time after their virialization. The
size of the oscillaton depends on the central value of the scalar
field, so that using (\ref{Mcrit}) and  the numerical results of
\cite{chipocludo}, we find that if the scalar field mass is
$m_{\Phi}=1/n\times8.11\times10^{-11}$eV, where $n$ is a constant
factor, the associated oscillatons will have a citical mass and
maximal radius
\begin{equation}m_{\textrm{crit}} \ = \ n\,M_{\odot},\qquad
R_{\textrm{max}} \ \sim \ 10\,\times\,
n\,\,\textrm{km},\label{RM}\end{equation}
Thus, for an oscillaton of $50\,M_{\odot}$, the scalar field mass
is $m_{\Phi}=1.6\times 10^{-12}\,$eV and its size is $r_* \sim
R_{\textrm{max}}\sim 500\,\textrm{km}$, while an oscillaton with
the earth's mass will be just a few meters across.

Boson stars, the other type of scalar MACHOs, are static rather
than oscillating and exist in systems of more than one scalar
field \cite{Liddle1992,Jetzer1992,LeePang1992}. In their simplest
form they are described by a massive complex scalar field with a
possible self-interaction:
$U=m_{\Phi}^2|\Phi|^2/2+\lambda|\Phi|^4/4$. There are two
quantitatively different cases. When there is no self-interaction
($\lambda=0$), boson stars are formed with a mass of the order of
$m_{Pl}^2/m_{\Phi}$ and radius which is a little larger than the
corresponding Schwarzschild radius, as for oscillatons. The
critical mass is given by a relation similar to equation
(\ref{Mcrit}):
\begin{equation}
m_{\textrm{crit}} \ = \ 0.633 \frac{m_{Pl}^2}{m_{\Phi}} .
\label{McritBS}
\end{equation}
Self-interaction changes the situation dramatically since this
term dominates as long as $\lambda\gg ({m_{\Phi}}/{m_{Pl}})^2$
which holds almost for any value of $\lambda$. In that case the
boson star masses are much larger - roughly by a factor of
$\sqrt{\lambda} \, m_{Pl}/m_{\Phi}$. The critical
mass is given now by
\begin{equation}
m_{\textrm{crit}}  \approx \ 0.06\sqrt{\lambda} \,
\frac{m_{Pl}^3}{m_{\Phi}^2} . \label{McritBSSelfInt}
\end{equation}
Consequently, for a similar solar mass MACHO the scalar particle
mass should be taken now to be of the order of 1 GeV (for
$\lambda\sim 1$). As in the case $\lambda=0$, their radii are still of
the order of the corresponding Schwarzschild radius. Thus, the relation
between critical mass and maximal radius given in (\ref{RM}) is valid
(to a good approximation) also for boson stars. If we assume a
boson star formation in the early universe, starting after the temperature
drops below $m_{\Phi}$, there is a strong preference for boson star
formation by the self-interacting type of bosons over formation by free
bosons \cite{Liddle1992,Jetzer1992}. Most of the bosons of the latter
type tend to condensate into black holes rather to form boson
stars while the opposite is true for the former. 

In either case, boson stars or oscillatons, the idea of each halo
being a single solitonic configuration runs into some problems:
once fundamental scalar field parameters are chosen, the size of
the ``boson star halo'' is fixed \cite{bosonS} so that halos of
only one unique size would exist, in stark disagreement with
observations. Regarding the case of ``single oscillaton halos'',
their oscillations would probably lead to halo--scale astronomical
effects that should have been detected. 

It has also been
suggested~\cite{TK1,TK2} that ``axitons'', \textit{i.e.} boson star MACHOs in
the form of sub planet--size solitons ($m<10^{-9}\,M_\odot$) of the axion field  
possibly produced around the QCD epoch of the universe, might account for a large
proportion (but not all) of non--baryonic dark matter in galactic halos. This
type of ``partial'' approach (see also~\cite{MielkeS}) can not be ruled out, but
does not significantly add  to the discussion beyond further introducing an extra
free parameter, as the  remaining fraction of the halo mass still has to be
accouned for through a separate type of dark matter. As long as a unique type of
dark matter at  all galactic and cosmological scales remains feasible, we believe
it is  this the framework within which one must work, on grounds of conceptual 
economy.

\section{Halos made up of scalar machos}

In this paper we propose an alternative approach in which each
galactic dark matter halo is not a single halo--sized
oscillaton or boson star, but is a collisionless ensemble of such
objects: {\textit{i.e.}} ``scalar field MACHOs''. In other words,
we consider a scenario in which the scalar field evolves by
forming a large number of stable star--sized or planet--sized scalar
condensations which end up clustering into structures similar to
standard CDM halos (but made of scalar field MACHOs instead of WIMPs). We assume
further that these scalar MACHOs constitute the totality of non--baryonic dark
matter.

Under the
proposed scheme, these dark matter halos would follow very similar dynamics to the
usual CDM halos, but with a different particle mass granularity
given by the microlensing and dynamical constraints on the MACHOs's masses. Note
that numerial N-body simulations at ultra high resolution have shown results
converging for particle numbers higher than $4 \times 10^7$ for Milky Way type
halos~\cite{ghigna}, {\it i.e.} they are insensitive to particle mass granularity
smaller than around $10^5 M_\odot$. Therefore, from a dynamical and
phenomenological point of view, cosmological N-body simulations modeling CDM
particles would be indistinguishable from those modeling scalar MACHOs with $m\alt
10^5 M_\odot$.  

In order to be consistent with the MACHO detection
constraints from microlensing~\cite{Microlensing}, the masses of the
scalar MACHOs making up the galactic halos must lie in the ranges: \, $ m \agt
30\,M_{\odot}$ \,or\, $ m\alt 10^{-7}\,
M_{\odot}$. An upper bound on the scalar MACHO mass also arises from various
considerations. First, even the smallest dwarf galactic halo
with $M\sim 10^8\, M_{\odot}$ must contain at least $\sim 1000$ scalar
MACHOs so that it can be modelled as a stable colissionless ensemble of
these objects, hence we should have $m\alt 10^5\,
M_{\odot}$. Secondly, since the scalar MACHOs we
are considering are very compact objects, we can apply to them the
arguments used in various articles that examine the proposal
galactic halos are made up of super-massive black holes (BH)~\cite{mBH1}. These
papers argue that BH's larger than $10^3\,M_\odot$ would lead to a central
BH that is too large \cite{mBH2,mBH3,mBH4} or would destroy the observed globular
cluster halo population through dynamical interactions
\cite{mBH5,mBH6,mBH7}. Considering these arguments plus the 
microlensing constraints, the allowed mass range for the MACHO scalars must be
\begin{equation}  m\alt 10^{-7}\,
M_{\odot} \quad
\textrm{or}\quad 30\,M_{\odot}\alt m\alt
1000\,M_\odot.\label{massranges}\end{equation}
Notice that in the lower end of this range there is no
inherent minimal bound on the MACHO mass. We can think of planetary or asteroid 
size ``mini--MACHOs'' or, in principle, even much smaller ones, as long as they can
be described as an ensemble of classical particles. Extremely
small scalar mini--MACHOs can also be described as some sort of very
large scalar field WIMPs.

\section{Gravothermal instability}
 
Observations 
from LSB and  dwarf galaxies, overwhelmingly dominated by dark matter, seem to
exhibit flat constant density cores \cite{vera, LSB}, instead of the high density
core surrounded by a lower density halo characteristic of the gravothermal
instability. Hence, halos made up of ensambles of scalar MACHOs must be
characterized by two--body relaxation timescales larger than
$13.7$\, Gyr, the estimated age of the universe (according to latest WMAP
estimates~\cite{MAP}). This relaxation timescale is~\cite{BT}. 
\begin{equation}\tbr \ = \ \frac{1.8\times
10^{10}\,\textrm{yr}}{\ln
\,(0.4\,M/m)}\,\frac{M_\odot}{m}\,\frac{M_\odot\,\textrm{pc}^{-3}}
{\rho}\,\left(\frac{\sigma}{\textrm{km}\,\,\textrm{s}^{-1}}
\right)^3,\label{trel}\end{equation}
where $\sigma$ is the velocity dispersion, $\rho$ is mass density and $M$ is the
halo mass (the virial mass), so that $M/m=(1/n)\,(M/M_\odot)$ for a scalar MACHO
mass $m=\,n\,M_\odot$. Notice that $\tbr$ varies from point to point along the
halo, up to several orders of magnitude between the center and outlaying regions.
As it happens with globular clusters and in numerical simulations, the
highest density central region might be older than $\tbr$, but not the
outlying lower density regions. However, if the center region is younger
than $\tbr$, so will the rest. Therefore, it is
sufficient to evaluate (\ref{trel}) for the central values $\rho=\rho_c$
and $\sigma=\sigma_c$ in order to provide a criterion for a gas of scalar
MACHOs \textit{not} to be older than $\tbr$.  This
criterion is
\begin{equation} \tbr|_c \  >
\ 13.7\,\textrm{Gyr}\label{condtrel}\end{equation}
where $\tbr|_c$ is (\ref{trel}) evaluated for $\rho=\rho_c$
and $\sigma=\sigma_c$. Since $\sigma_c$ is related to the maximal rotation
velocity, it can be thought of as the characteristic scale parameter of
different halo structures, hence for same sized halos the relaxation state
depends mostly on $\rho$ and $m$.
\begin{figure}
\centering
\includegraphics[height=6cm]{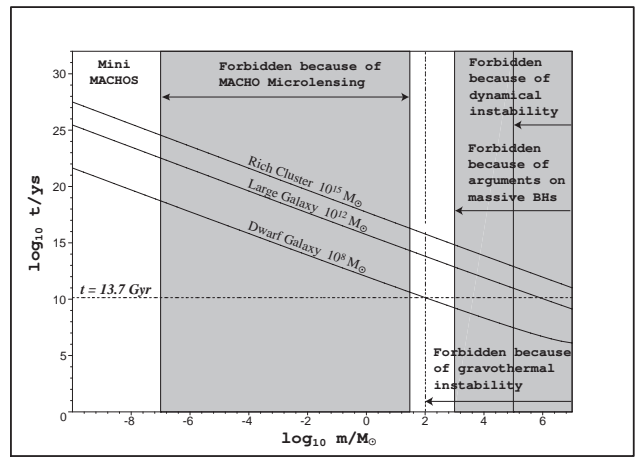}
\caption{The central two--body relaxation time in years for
various galactic halo structures as a function of the scalar MACHO mass (as
fraction of $M_\odot$). }
\label{fig:trel_new2}
\end{figure}

We examine in figure \ref{fig:trel_new2} the fulfilment of
(\ref{condtrel}) for both types of scalar MACHO candidates, by
plotting $t_{\textrm{rel}}|_c$ vs the scalar MACHO mass $m$ for
various typical halo structures characterized by specific values
of $\sigma_c$, that is: $10\, \textrm{km}\,\textrm{s}^{-1}, 200\,
\textrm{km} \,\textrm{s}^{-1}$ and $1000\,\textrm{km}
\,\textrm{s}^{-1}$, for a dwarf galaxy, a large galaxy and a
galaxy cluster, with $M=10\,^8,\,10^{12},\,10^{15}\,\,M_\odot$,
respectively. We use $\rho_c=
1\,M_\odot\,\textrm{pc}^{-3}$, the largest in the range of
estimated values~\cite{ranges}
\begin{equation} 0.001 \,\, M_\odot\,\textrm{pc}^{-3} \ \alt \ \rho_c
\ \alt \ 1\,\, M_\odot\,\textrm{pc}^{-3},\label{rhoranges}\end{equation}
since, if the halo has not
undergone core collapse for this value, it will not for smaller
values in the range (\ref{rhoranges}).  The shaded regions are the mass
ranges excluded in (\ref{massranges})  and the
horizontal dashed line gives the current age of the universe, a
good estimate of the lifetimes of the oldest of these systems. Hence, the most
stringent test comes from verifying (\ref{condtrel}) for the
smallest/densest halos. As shown by figure \ref{fig:trel_new2}, this
condition is not fulfilled for scalar MACHOs with
mass $m\agt 100\,M_\odot$. Since we would expect all galactic
halos to be made of the same type of scalar MACHOs (same type of
dark matter), figure \ref{fig:trel_new2} would be indicating that
for scalar MACHOs with $m\agt 100\,M_\odot$ core--collapse would
happen in small halos but not in larger structures. However, this
scenario is at  odds with observations in LSB galaxies of various
sizes~\cite{vera,LSB},
all of which exhibit approximately isothermal density profiles.

\section{conclusion}

From the discussion above, the allowed masses must lie in
the mini--MACHO range: $m\alt 10^{-7}\,M_\odot$, and within
$30\,M_\odot\alt m \alt 100\,M_\odot$. These values roughly 
agree with the findings of Yoo {\it et al}~\cite{mBH8}, who rule out massive BH's
with $m> 43\,M_\odot$ making up our galactic halo, as they would result in a
depletion of halo binaries that contradicts observations. Since the window of
acceptable masses for the large scalar MACHOs is rather narrow, our results
and those of \cite{mBH8} do not suggest ``the end of the MACHO era'' (as claimed
in \cite{mBH8}), but that MACHOs making up dark matter halos are, most probably,
scalar field mini--MACHOs.  This proposal not only solves the problems of scalar
field dark matter, but does not violate the constraints arising from Big Bang
Nucleosynthesis because the scalar field is not baryonic. Furthermore, at
cosmological scales the proposed ensembles of scalar MACHOs behave just as CDM,
though a non-supersymmetric type of CDM. We believe that this alternative modeling
can describe the early universe origin of dark matter by means of new fundamental
physics, while at the same time keeping the advantages of the phenomenology of the
thermal CDM paradigm. A more detailed and less idealized study of the proposed
dark matter scenario is presently under consideration.

\acknowledgements

This work was partly
supported by CONACyT M\'exico (grants 32138-E and 34407-E) and PAPIIT--DGAPA
(grant IN117803).

\end{document}